\documentstyle[preprint,aps,eqsecnum]{revtex}
\begin{document}
\draft

\tightenlines
\newcommand{\modegy}{\,({\rm mod}\,1)}
\newcommand{\modketto}{\,({\rm mod}\,2)}
\newcommand{\J}{{\cal J}}
\newcommand{\ch}{{\rm ch}}
\newcommand{\sh}{{\rm sh}}
\newcommand{\th}{{\rm th}}
\newcommand{\bml}{\begin{mathletters}}
\newcommand{\eml}{\end{mathletters}}
\newcommand{\be}{\begin{equation}}
\newcommand{\ee}{\end{equation}}
\newcommand{\ba}{\begin{array}}
\newcommand{\ea}{\end{array}}
\newcommand{\bea}{\begin{eqnarray}}
\newcommand{\eea}{\end{eqnarray}}
\newcommand{\eps}{\epsilon}
\newcommand{\pa}{\partial}
\newcommand{\lb}{\lbrack}
\newcommand{\rb}{\rbrack}
\newcommand{\de}{\delta}
\newcommand{\ka}{\kappa}
\newcommand{\h}{{1\over2}}
\newcommand\delsl{\raise.15ex\hbox{/}\kern-.57em\partial}
\newcommand{\bm}{B^-}
\newcommand{\bp}{B^+}

\def\eqalign#1{
\null \,\vcenter {\openup \jot \ialign {\strut \hfil $\displaystyle {
##}$&$\displaystyle {{}##}$\hfil \crcr #1\crcr }}\,}

\title{Scaling limit of the one-dimensional attractive  
Hubbard model: The non-half-filled band case}

\author{F. Woynarovich}
\address{Institute for Solid State Physics\\
of the Hungarian Academy of Sciences\\
1525 Budapest 114, Pf 49.}
\author{P. Forg\'acs}
\address{Laboratoire de Math.~et Physique Th\'eorique\\
CNRS UPRES-A 6083\\
D\'epartement de Physique\\
Facult\'e des Sciences, Universit\'e de Tours\\
Parc de Grandmont, F-37200 Tours}
\maketitle
\begin{abstract}
The scaling limit of the less than half filled attractive Hubbard chain is 
studied. This is a continuum limit in which the particle number per lattice
site, $n$, is kept finite ($0<n<1$) while adjusting 
the interaction and bandwidth
in a such way that there is a finite mass gap.
We construct this limit both for the spectrum and the secular equations
describing the excitations. We find, that similarly to the half filled
case, the limiting model has a massive and a massless sector.
The structure of the massive sector is closely analogous to that of the
half filled band and consequently to the chiral invariant SU(2) Gross-Neveu 
(CGN) model. The 
structure of the massless sector differs from that of the half filled 
band case: the excitations are of particle and hole type, however
they are not uniquely defined. 
The energy and the momentum of this sector exhibits a tower
structure  
corresponding to a conformal field theory with $c=1$ and 
SU(2)$\times$SU(2) symmetry. 
The energy-momentum spectrum and the 
zero temperature free energy of the states with finite density
coincides with that of the half filled case supporting
the identification of the limiting model with the 
SU(2) symmetric CGN theory.
\end{abstract}


\section{INTRODUCTION}
\label{sec:int}

The one dimensional (1D) Hubbard model is described by the Hamiltonian
\bea\label{1}
\hat H=&-&t\sum_{i=1}^N\sum_{\sigma=\uparrow,\downarrow}
  \left( c^{+}_{i,\sigma}c^{\phantom{+}}_{i+1,\sigma}+h.c.\right) +
  U\sum_{i=1}^N\left( \hat n_{i,\uparrow}-{1\over2}\right)
  \left( \hat n_{i,\downarrow}-{1\over2}\right)\nonumber\\
  &+&\mu\sum_{i=1}^N\left( \hat n_{i,\uparrow} + \hat n_{i,\downarrow}\right)
\eea
in which $c^{+}_{i,\sigma}(c^{\phantom{+}}_{i,\sigma})$ 
creates (destroys) an electron at 
the site $i$ with spin $\sigma$, 
$\hat n_{i,\sigma}=c^{+}_{i,\sigma}c^{\phantom{+}}_{i,\sigma}$.
We impose periodic boundary conditions so the site $i=N+1$ is 
the equivalent to the site $i=1$. The hopping $t$ is positive while the 
interaction $U$, for the considered attractive case, is negative, 
and the value of the 
chemical potential, $\mu$, is choosen to fix the  
particle number per site, $n$. This model being completely integrable 
\cite{Sha}
and exactly solvable by the Bethe Ansatz (BA) \cite{LiWu} plays a central
role in the theory of strongly correlated electron systems \cite{KoEss}. 
At the same
time by direct linearization around the Fermi-points \cite{So}
both the {\it half-filled} ($n=1$) and the {\it non-half-filled} 
($n<1$) Hubbard chains
can be related to relativistic field theory models 
\cite{GrNe,WiLa}, specifically to the SU(2) symmetric chiral
Gross-Neveu (CGN) model. Since the Hamiltonian of the CGN model has been 
diagonalized  in the continuum (with a somewhat unorthodox cutoff i.e.\ 
filling up the Dirac sea up to a certain depth) by the BA method
\cite{AnLo}, constructing the 
relativistic (scaling) limit of the
BA {\it solution} of the Hubbard chain can be of twofold interest.
On the one hand it makes possible to study 
the 
relation between the two models in the detail,
and on the other hand it promotes
the Hubbard chain to an {\sl integrable lattice} regularization of
the CGN model.

Earlier Filev \cite{Fi}, later Melzer \cite{Me}, 
recently Woynarovich and Forg\'acs \cite{WoFo} have studied the 
relativistic limit of the {\it half-filled} (HF) Hubbard chain. 
In ref.\ \cite{WoFo} we have given a rather complete
study of the structure of the states and the spectrum of the limiting model 
and also presented some convincing arguments that the scaling limit of the
HF attractive Hubbard chain is 
the SU(2) (in fact SO(4)) symmetric CGN model. 
The scaling limit of the {\it non-half-filled} (NHF)
attractive Hubbard chain
has recently been studied by Woynarovich \cite{Wo0}. He has shown, that
the limiting model possesses a massive sector whose spectrum and phaseshifts
agree with those found in the scaling limit of the HF Hubbard chain. 

The aim of the present work is to extend the study of Ref.\ \cite{Wo0}
to the complete spectrum and to collect some other arguments supporting
the equivalence of the scaling limit of the {{\it non-half-filled} and that of 
the {{\it half-filled} Hubbard chain. This also implies that the scaling limit of
the {{\it non-half-filled} Hubbard chain is the SO(4) symmetric CGN
model.

The scaling limit is a continuum limit $N\to\infty$, $a\to0$ so that
$Na=L={\rm const.}$ ($a$ and $L$ being the lattice constant and the 
chain length, respectively), in which the particle number ($N_e$) per 
site $N_e/N\to n$ 
finite, i.e.~$N_e\to\infty$ too. To avoid divergences,
the interaction $u$ has to 
be tuned in a special way (actually $u\to 0$). Finally, although $a$ 
does not appear explicitly in $\hat H$, since $t\propto1/{\rm distance}$, 
$t\to\infty$ as $a\to0$. All this is to be performed so that the gap in the
spectrum of the unbound electrons is kept finite, that is \cite{Wo0}:
\bml\label{scl}
\bea\label{scl.a}
&&u\to0,\quad t\to\infty;\quad N,N_e\to\infty,\quad{\rm at}
\quad N_e/N\to n={\rm const}<1;\nonumber\\
&&a\to0\quad{\rm at}\quad Na=L={\rm const}
\eea
so that
\be\label{scl.b}
   m_0={8t\over\pi}\sqrt{u\sin^3(\pi n/2)}\,
   {\rm exp}\left\{-{\pi\sin(\pi n/2)\over2u}\right\}={\rm const}
\ee
and
\be\label{scl.c}
 2at\sin(\pi n/2)=1\ ,
\ee
\eml
We have performed this limit both in the spectrum and in the higher level
Bethe Ansatz (HLBA) equations of the NHF Hubbard chain. We can summarise
the properties of the limiting theory as follows:
\begin{itemize}
\item[(i)] Like in the HF case, there are two kinds of excitations:
massive and massless ones. 
While the massive sector is
described in terms of well defined particles, 
in the massless sector the definition of the particless is not unique.
(This is a major difference as compared to the HF case, where the excitations
in both sectors are well defined particles.) 
\item[(ii)] The massive particles have spin 1/2. Their 
contribution to the energy and momentum is given as
\bml\label{9}
\be\label{9.a}
  \sum_{j}\epsilon(\kappa_j),\quad{\rm and}\quad 
  \sum_{j}p(\kappa_j),
\ee
where 
\be
  \epsilon(\kappa)=m_0\cosh(\kappa),\quad\quad 
  p(\kappa)=m_0\sinh(\kappa),
\ee
\eml
and $\kappa_j$'s are the rapidities.
The $\kappa_j$'s and the set of variables 
$\chi_{\alpha}$ describing the spin state of the particles,
satisfy the following BA type equations
\bml\label{10}
\be\label{10.a}
  Lp(\kappa_j)=2\pi I_{j}-\sum_{j'}
  \phi\left({\kappa_j-\kappa_{j'}\over\pi}\right)+
  \sum_{\alpha}2{\rm tan}^{-1}\left({\kappa_j-\chi_{\alpha}\over\pi/2}\right)\ ,
\ee
\bea\label{10.b}
  &&\sum_j 2{\rm tan}^{-1}\left({\chi_{\alpha}-\kappa_j\over\pi/2}\right)=
  2\pi J_{\alpha}+
  \sum_{\alpha'}2{\rm tan}^{-1}
  \left({\chi_{\alpha}-\chi_{\alpha'}\over\pi}\right)\ ,
\eea                             
\eml  
with
\be\label{fi}  
  \phi(x)={1\over i}{\rm ln}{\Gamma\left({1\over2}-i{x\over2}\right)
                             \Gamma\left(1+i{x\over2}\right)\over
                             \Gamma\left({1\over2}+i{x\over2}\right)
                             \Gamma\left(1-i{x\over2}\right)}\ .
\ee
(\ref{10}a,b) yield  
the same phaseshifts as those obtained for the 
half-filled band \cite{Me}
\be\label{11}
  \psi^{tr}=\pi+\phi\left({\Delta\kappa\over\pi}\right),\quad
  \psi^{s}=\phi\left({\Delta\kappa\over\pi}\right)-
  2\tan^{-1}\left({\Delta\kappa\over\pi}\right)\quad
  (\Delta\kappa=\kappa_j-\kappa_{j'}).
\ee
\item[(iii)] The massless excitations correspond to states above resp.\ 
below a certain Fermi niveau (particles resp.\ holes).
As the choice of the Fermi niveau is not
unique, however, the parametrization of the states (number of particles and holes,
and their momenta) is not uniquely defined. The energy and momentum, 
unlike in the HF case, consist not only of the sums 
of contributions of the individual excitations,  
but also contain certain `collective' terms too. The spectrum of this
sector shows the tower structure characteristic of a conformal field theory 
(CFT):
\bea
E-E_m&=&-{\pi\over6L}+{2\pi\over L}(x_{n,m}+\nu^++\nu^-)\,,\nonumber\\
P-P_m&=&{2\pi\over L}(s_{n,m}+\nu^+-\nu^-)\,,
\eea
where $E_m$ and $P_m$ being the energy and momentum of the vacuum plus
the massive sector, $\nu^{\pm}$ are integers and
\be
x_{n,m}={1\over2}\left(n^2+m^2\right)\,,\quad s_{n,m}=nm\,.
\ee
Here the numbers $n$ and $m$ are integers or half integers depending
on the parity of the number of massive particles.
In the case of an empty massive sector the apexes of the towers 
yield the conformal weights (the notations are those of \cite{Car})   
\be
\Delta={1\over4}(n+m)^2\,,\quad\bar\Delta={1\over4}(n-m)^2\,.
\ee
coinciding with those of a $c=1$ CFT with an (enhanced) SU(2) symmetry.
\item[(iv)] The terms corresponding to the 
apexes of the towers in the spectrum of the massless
sector depend nonlinearly on the number of the various excitations,
hence we refer to them as collective terms. The parametrization
of the massless sector (the Fermi niveau), however, can be choosen so, that
the number of massive particles formally disappears from the
collective terms and then the two sectors practically decouple.  
\item[(v)] 
The energy-momentum spectrum of the limiting model
is the same as that of the HF band case: the contribution of the 
massless sector as a function of the parameters $m$ and $n$ coincides
with  
the corresponding contribution found in Ref.\ \cite{WoFo}, and so does
the contribution of the massive sector too, as
the parity prescription
for the $I_j$ quantum numbers (i.e.\ the quantization of the momenta) expressed 
in terms of the $m$ and $n$ quantum numbers is also the same 
as that found in Ref.\ \cite{WoFo}.
 
\item[(vi)]
The ground state energy of states with a finite density of excitations  
is also in agreement with the result for the HF case. 
If we choose the Fermi surface so, that the two sectors decouple,
the zero temperature free energies
(which are nothing but the ground state
energies in the presence of a chemical potential)
of the different sectors become independent 
of each other, and the total free energy $f(\mu,\nu)$ can be written
as 
\be\label{1.10}
f(\mu,\nu)=-{\mu^2\over\pi} \Psi({\mu\over m_0})-{\nu^2\over4\pi}\,,
\ee
where $\mu$ and $\nu$ denote the chemical potentials for the 
massive resp.\ massless particles. The $\Psi({\mu/ m_0})$
can be given as an asymptotic series
of $1/[\ln(\mu/m_0)]$ for high particle densities.
(We remark that by parametrizing the massless sector in a different way,
other terms depending nonlinearly on both chemical potentials would appear in
Eq.\ (\ref{1.10}).)
\end{itemize}

The equivalence of the limiting models of the HF and NHF cases is not
trivial at all. While in a finite length chain we see a smooth behaviour
as the bandfilling $n$ is changed, in the thermodynamic limit the HF and
NHF cases separate. There are three major (although not independent) differences.
\begin{itemize}
\item[--] While the HF chain has two SU(2) symmetries 
(spin and isospin, this latter being connected to the charge), 
the NHF has only one. 
\item[--]
The structures 
of the excitation spectra are different, as in the HF case the gapless 
excitations are 
isospin 1/2 particles, while in the NHF case these excitations are particle
and hole type. 
\item[--] In the 
(naive) continuum limit of the HF case there are umklapp processes
violating chiral symmetry, while no such terms
are present in the NHF case.
\end{itemize}
It is somewhat surprising, that some of these differences 
thought to be significant, disappear in the
scaling limit: as we discussed in \cite{WoFo} there are indications, 
that the amplitude of the umklapp processes for the NHF chain scale out 
in a renormalization 
process, and as the conformal weights show, the SU(2) symmetry of the
massless sector of the NHF chain developes in the scaling limit. The only 
significant difference we see after the scaling limit is in the 
structure of the massless sector. On the other hand, the fact, that the 
energy-momentum
spectra of the two limiting theories coincide is a strong 
evidence supporting 
the equivalence of the two theories.  

The paper is organized as follows. In Sec.\ \ref{sec:bae} we give the BA 
equations of the Hubbard chain, and in Sec.\ \ref{sec:snhf} we describe
the solutions corresponding to the less than half filled band. 
The scaling limit of the secular equations and the spectrum are constructed 
in Secs.\ \ref{sec:eqscl} and \ref{sec:spscl} respectively, and the results 
are discussed in Sec.\ \ref{sec:interpr} together 
with a brief review of the states with a finite 
density of excitations, and we list the differences
between the limiting models of the HF and NHF chains in 
Sec.\ \ref{sec:diffs}. Appendix \ref{sec:nlim} we construct the naive 
continuum limit of the model.  

\section{THE BA EQUATIONS}
\label{sec:bae}

The eigenvalue equations of the 
Hamiltonian (\ref{1}) have been reduced to a set of nonlinear equations by Lieb and 
Wu \cite{LiWu}:
\bml
\bea\label{2.a}
 Nk_j&=&2\pi I_j-\sum_{\alpha = 1}^M2\tan^{-1}{\sin k_j-\lambda_{\alpha}
 \over U/4}\ ,\\
 \biggl(I_j&=&{M\over2}{\modegy}\biggr)\ ,\nonumber
\eea
\bea\label{2.b}
 \sum_{j=1}^{N_e}2\tan^{-1}{\lambda_{\alpha}-\sin k_j\over U/4}&=&
 2\pi J_{\alpha}+ \sum_{\beta=1}^M 2\tan^{-1}{\lambda_{\alpha}-
 \lambda_{\beta}\over U/2}\ .\\
 \biggl(J_{\alpha}&=&{N_e+M+1\over2}{\modegy}\biggr)\ .\nonumber
\eea
\eml
Here $N_e$ is the number of electrons, $M$ is the number of down spins, i.e. 
$S^z=(N_e/2-M)$, 
and the $I_j$ and $J_{\alpha}$ quantum numbers are integers or 
half-odd-integers depending on the parities of $N_e$ and $M$, as indicated. 
Once these equations are solved the wave-function can be given \cite{Wo1}
and also the 
energy and the momentum of the corresponding state can be calculated:
\be\label{3}
 E={NU/4}-\sum_{j=1}^{N_e}(2t\cos k_j+U/2-\mu),\quad\quad
 P=\sum_{j=1}^{N_e}k_j\ .
\ee
For the  considered $U<0$ attractive chain near the ground-state 
most of the electrons
form bound pairs with wavenumbers given (up to corrections exponentially 
small in $N$) as  
\be\label{4}
 \sin k^{\pm}=\Lambda\pm iu
\ee
with $u=|U|/4t$ and ${\Lambda}$ being a subset of the set ${\lambda}$. 
By this relation $k^{\pm}$ can be eliminated from Eq.(\ref{2.b})
and one finds that the wavenumbers of the unbound electrons, the $\lambda$s
connected with their spin distribution and the $\Lambda$s of the bound pairs
satisfy the equations \cite{Wo2,WoPe}
\bml\label{5}
\bea\label{5.a}
 2\pi I_j&=&Nk_j - \nonumber\\
 &&-\sum_{\alpha = 1}^{n(\lambda)}
 \varphi_1(\sin k_j-\lambda_{\alpha})-
 \sum_{\eta = 1}^{n(\Lambda)}
 \varphi_1(\sin k_j-\Lambda_{\eta})\ ,\\
 \biggl(I_j&=&{n(\lambda)+n(\Lambda)\over2}{\modegy}\biggr)\nonumber
\eea
\bea\label{5.b}
\sum_{j=1}^{n(k)}\varphi_1(\lambda_{\alpha}-\sin k_j)&=&
 2\pi J_{\alpha}+ \sum_{\beta=1}^{n(\lambda)}\varphi_2(\lambda_{\alpha}-
 \lambda_{\beta})\ ,\\
 \biggl(J_{\alpha}&=&{n(k)+n(\lambda)+1\over2}{\modegy}\biggr) \nonumber
\eea
\eml
\bea\label{6}
  2\pi J_{\eta}&=&N\left(\sin^{-1}\!(\Lambda_{\eta}-iu)+
  \sin^{-1}\!(\Lambda_{\eta}+iu)\right)\,-\nonumber\\
  &-&\sum_{j=1}^{n(k)}\varphi_1(\Lambda_{\eta}-\sin k_j)
  -\sum_{\nu=1}^{n(\Lambda)}\varphi_2(\Lambda_{\eta}-
  \Lambda_{\nu})\ .\\
  \biggl(J_{\eta}&=&{n(k)+n(\Lambda)+1\over2}{\modegy}\biggr)\nonumber
\eea
with
\be
  \varphi_m(\xi)=2\tan^{-1}(\xi/um)\label{varphim}
\ee
Here $n(k)$, $n(\lambda)$ and $n(\Lambda)$ are the number of unbound electrons,
the number of unbound electrons with down spins and the number of bound pairs,
respectively ($N_e=n(k)+2n(\Lambda)$, $M=n(\lambda)+n(\Lambda)$), and the 
quantum numbers 
$I_j$, 
$J_{\alpha}$
and $J_{\eta}$ are integers or half-integers as indicated. (As in  the 
following manipulations the quantum numbers may be redefined by absorbing 
certain constants into them, we give the `parity prescriptions' for the 
quantum numbers together with the equations. All these prescriptions hold 
for even $N$) 
The energy and momentum expressed by these variables is
\bea
  E=&&-Nut-\sum_j \left( 2t(\cos k_j-u)-\mu\right)-\nonumber\\
  &&-\sum_{\eta}\left(2t\left(\sqrt{1-(\Lambda_{\eta}-iu)^2}
  +\sqrt{1-(\Lambda_{\eta}+iu)^2}-2u\right)-2\mu\right)\ ,\label{energy}\\
  P=&&{2\pi\over N}\left(\sum_jI_j-\sum_{\alpha}J_{\alpha}+\sum_{\eta}J_{\eta}
  \right)\label{momentum}\ .
\eea
\section{STATES OF THE NON-HALF-FILLED HUBBARD CHAIN}
\label{sec:snhf}

\centerline{\it 1. Ground state -- the physical vacuum}
\medskip
As a reference state consider the state, consisting of  
$N_r=n_r(\Lambda)<N/2$ 
bound pairs (and no $k$'s, i.e.\ $N_e=2N_r$), and 
the quantum number set $J_{\eta}$ is given by 
\be
  J_1=-{N_r-1\over2},\quad\quad J_{\eta+1}=J_{\eta}+1,\quad\quad
  J_{N_r}={N_r-1\over2}.
\ee
If both $N,N_r\to\infty$ (at $N_r/N=n/2$ kept constant) the 
$\Lambda_{\eta}$'s
will be distributed in an interval $-B<\Lambda<B$ with a density $\sigma_r$
that is defined so, that the number of $\Lambda_{\eta}$s within the interval
$(\Lambda,\Lambda+d\Lambda)$ is given as $N\sigma_r(\Lambda)d\Lambda$. 
It is not hard to see from (\ref{6}), that this density is the derivative 
of the so called
{\it counting function}:
\be\label{szigmar}
\sigma_r(\Lambda)={dz_r(\Lambda)\over d\Lambda}\ ,
\ee
where
\be\label{zer}
  z_r(\Lambda)={1\over2\pi}\left\{\left(\sin^{-1}\!(\Lambda-iu)+
  \sin^{-1}\!(\Lambda+iu)\right)\,-{1\over N}\sum_{\eta=1}^{N_r}
  \varphi_2(\Lambda-
  \Lambda_{\eta})\right\}\ .
\ee
This relation, applying the leading term of the Euler-Maclaurin
type summation formula
\be\label{EuMac}
{1\over N}\sum_{\eta=1}^{n}f\left(J_{\eta}\over N\right)=
\int\limits_{J_1/N-1/2N}^{J_n/N+1/2N}\!\!\!f\left(x\right)dx-
{1\over24N^2}\left(f'\left(J_n+1/2\over N\right)-
f'\left(J_1-1/2\over N\right)\right)
\ee
leads to the equation
\bea
\sigma_r(\Lambda)=&&\sigma_0(\Lambda)-{1\over2\pi}\int\limits_{-B}^{B}
  K_2(\Lambda-\Lambda')\sigma_{r}(\Lambda')\ ,\label{szigmarint}\\
  &&\sigma_0(\Lambda)=
  {1\over2\pi}2{\rm Re}\left(\left(\sqrt{1-(\Lambda-iu)^2}\right)^{-1}
  \right)\ ,\\
  &&K_m(\xi)={2mu\over (mu)^2+{\xi}^2}\,,\label{K_m}
\eea
where the limits are determined through     
\bml\label{B}
\be
z_r(\pm B)=\pm{N_r/2N}
\ee
(i.e.
\be\label{Be}
  \int\limits_{B}^{\infty}\sigma_{r}(\Lambda)=
  \int\limits^{-B}_{-\infty}\sigma_{r}(\Lambda)=
  {1\over2}\left(1-{2N_r\over N}\right)={1-n\over2}\ ,
\ee
\eml
as it can be seen by integrating (\ref{szigmarint})).

The energy of this reference state is (using summation formula (\ref{EuMac})
in leading order)
\bea\label{Er}
  E_r=&&-utN+\sum\varepsilon_0(\Lambda_{\eta})\nonumber\\
  =&&-utN+N\int\limits_{-B}^{+B}\varepsilon_0(\Lambda)\sigma_r(\Lambda)\ ,
\eea
with
\be\label{eps0}
\varepsilon_0(\Lambda)=
  -\left(4t\left({\rm Re}\sqrt{1-(\Lambda-iu)^2}-u\right)
  -2\mu\right)\ .
\ee
In (\ref{Er}) the first term is irrelevant. 
The second term, iterating $\sigma_r$,
can be transformed into the form 
\be\label{EBmu}
E_r+utN=
  N\int\limits_{-B}^{+B}\varepsilon_r(\Lambda)\sigma_0(\Lambda)\ ,
\ee
where $\varepsilon_r$ satisfies the equation
\be\label{epsilonr}
\varepsilon_r(\Lambda)=\varepsilon_0(\Lambda)-
  {1\over2\pi}\int\limits_{-B}^{B}
  K_2(\Lambda-\Lambda')\varepsilon_{r}(\Lambda')\ .
\ee

It is easy to check, that the energy of (\ref{EBmu}) as a function of $B$,
i.e.\ as a function of $n$, (at fixed $\mu$) 
is minimal if 
\be\label{epsB=0}
\varepsilon_r(B)=0\,.
\ee
As $\varepsilon_0$, depends on $\mu$ linearly so does $\varepsilon_r$,
hence it is possible to choose a $\mu$ such that 
(\ref{epsB=0}) holds at the desired bandfilling $n$.
In the following we suppose that $\mu$ has this value, i.e.\ the
reference state is the ground state.

In general the (\ref{szigmarint}) can not be solved analytically, and $B$
as a function of $n$ can not be given in a closed form. In the relativistic
limit however, when $u\to0$, (\ref{szigmarint}) can be solved 
using the method devised by Yang and Yang \cite{YaYa}: the equations of the 
type (\ref{szigmarint}) can be transformed into a series of Wiener-Hopf
equations which can be solved with sufficient accuracy
yielding \cite{WoPe}
\be\label{B(un)}
  B=\sin{\pi n\over2}-{u\over\pi}\left(1+\ln{\pi\cos^2{\pi n\over2}
  \sin{\pi n\over2}\over2u}\right).
\ee

\bigskip
\centerline{\it 2. The possible excitations}
\medskip
The ground state
can be excited by combinations of the following three "elementary"
excitations:
\begin{itemize}
\item[i)] introducing holes and particles in the $\Lambda$ distribution
by removing some $J_{\eta}$ from the ground-state set and introducing some
outside it,
\item[ii)] introducing complex $\Lambda$s,
\item[iii)] introducing unbound electrons (real $k$s and $\lambda$s).
\end{itemize}
\noindent The excitations of type i) with the proper choice of $\mu$ 
(described above) have a dispersion with no gap, 
while those of type ii) and iii) possess gaps. The scaling limit
(\ref{scl}) is constructed so, that the gap in the spectrum of the excitations 
of type iii) is a fixed finite value ($m_0$). In this limit the gap of
the excitations of type ii) diverges, i.e.~they can be 
discarded. (In the following we shall refer to type i) excitations as
particles and holes or massless excitations, 
and type iii) excitations will be called massive 
particles or unbound electrons.)

\bigskip
\centerline{\it 3. Particles and holes in the $\Lambda$ set}
\medskip
First we consider the equations of the $\Lambda$s (\ref{6}).
To describe the particle and hole type excitations 
properly we should define the `Fermi sea'.
If $n(k)+n(\Lambda)=n_r(\Lambda)\modketto$  
the Fermi sea (of $J$s)
is the ground state set of the $J$s, i.e.~the Fermi sea is 
made up by all the $J_f$ integers or half-integers 
satisfying
\be\label{Fermisea}
J_1=-J_{F}+1/2\,,\quad\quad J_{f+1}=J_f+1\,,\quad\quad J_f\leq J_{F}-1/2\,,
\ee
with 
\be
J_F=N_r/2\,.
\ee
If $n(k)+n(\Lambda)\not=n_r(\Lambda)\modketto$, there is no unique choice:
as the parity of the $J_{\eta}$s (i.e.\ that of the $J_f$s) is changed, 
keeping (\ref{Fermisea}) we may
chose any of $J_F=(N_r\pm1)/2$. For the sake of definiteness we choose
\be
J_F=(N_r+1)/2\,,
\ee
thus
\be\label{delta}
J_F={N_r+\delta\over2}\,,\quad \delta=\left\{\begin{array}{ll}
0&\mbox{if $n(k)+n(\Lambda)=n_r(\Lambda)\modketto$}\\
1&\mbox{if $n(k)+n(\Lambda)\not=n_r(\Lambda)\modketto$}\end{array}\right.\,.
\ee
(We note that the Fermi sea defined above resembles most the 
ground-state distribution of the $\Lambda$s, nevertheless many other 
definitions are possible: the only role of
$J_F$ is to provide a reference point in the space of the $J$ quantum
numbers
connected to the $\Lambda$-rapidities, i.e.\ $\delta$ 
can be choosen even to depend
explicitly on $n(k)$. We shall discuss this later.)

To proceed, we define the counting function:
\bea
  z(\Lambda)&=&{1\over2\pi}\Biggl\{\left(\sin^{-1}\!(\Lambda_{\eta}-iu)+
  \sin^{-1}\!(\Lambda+iu)\right)\,-\nonumber\\
  &-&{1\over N}\sum_{j=1}^{n(k)}\varphi_1(\Lambda-\sin k_j)
  -{1\over N}\sum_{\eta=1}^{n(\Lambda)}\varphi_2(\Lambda-
  \Lambda_{\eta})\Biggr\}\ ,\label{ze}
\eea
in terms of which Eq.~(\ref{6}) reads
\be\label{Lambdakzevel}
  z(\Lambda_{\eta})={J_{\eta}\over N}\,.
\ee
It is clear, that (\ref{Lambdakzevel}) has a solution for any 
$z(-\infty)<J<z(+\infty)$ replacing $J_{\eta}$.  
The rapidities of the particles, holes and respectively the 
elements of the Fermi sea are determined by the equations
\be\label{Lambdak2}
z(\Lambda_p)={J_p\over N}\,,\quad\quad
z(\Lambda_h)={J_h\over N}\,,\quad\quad
z(\Lambda_f)={J_f\over N}\,,
\ee
respectively, where the $J_p$s are those $J_{\eta}$s, 
which are not elements of the set $J_f$ of (\ref{Fermisea}), while the 
$J_h$s are those $J_f$s, which are not elements of the $J_{\eta}$ set.
(Note, that $\pm J_F$ themselvs can be neither holes nor particles, as their
parity is different from that of the $J_{\eta}$s and $J_f$s.)
Comparing the parity prescription $J_{\eta}=(n(k)+n(\Lambda)+1)/2$ and the
number of elements in (\ref{Fermisea}) we see, that the total number
of $k$s (massive particles), particles and holes (massless excitations)
is always even. 
(It has to be noted, that the above mentioned freedom in choosing $J_F$ 
introduces an ambiguity in the 
{\it description} of certain states: depending on the
choice of $J_F$, in the same state the number
of massless excitations can be
different. This ambiguity drops out, however,
of physical quantities, like the energy and momentum which do not depend
on the choice of $J_F$.) The particles and holes being defined, it is convenient
to write $z(\Lambda)$ in the form
\bea\label{ze2}
  z(\Lambda)&=&{1\over2\pi}\Biggl\{\left(\sin^{-1}\!(\Lambda_{\eta}-iu)+
  \sin^{-1}\!(\Lambda+iu)\right)\,-\nonumber\\
  &-&{1\over N}\sum_{j=1}^{n(k)}\varphi_1(\Lambda-\sin k_j)
  -{1\over N}\sum_f\varphi_2(\Lambda-\Lambda_f)\nonumber\\
  &-&{1\over N}\sum_p\varphi_2(\Lambda-\Lambda_p)
  +{1\over N}\sum_h\varphi_2(\Lambda-\Lambda_h)\Biggr\}\,.
\eea
The density of the $\Lambda$s is given by the derivative 
of the $z$:
\be\label{sigmat0} 
\sigma_t(\Lambda)={dz(\Lambda)\over d\Lambda}\,,
\ee
i.e.
\bea\label{sigmat1}
  \sigma_t(\Lambda)&=&\sigma_0(\Lambda)
  -{1\over2\pi N}\sum_{j=1}^{n(k)}K_1(\Lambda-\sin k_j)
  -{1\over2\pi N}\sum_fK_2(\Lambda-\Lambda_f)\nonumber\\
  &-&{1\over2\pi N}\sum_pK_2(\Lambda-\Lambda_p)
  +{1\over2\pi N}\sum_hK_2(\Lambda-\Lambda_h)\,.
\eea
Calculating the sum over the Fermi sea by means of (\ref{EuMac})
we arrive at  
\bea\label{sigmat2}
  \sigma_t(\Lambda)&=&\sigma_0(\Lambda)
  -{1\over2\pi N}\sum_{j=1}^{n(k)}K_1(\Lambda-\sin k_j)\nonumber\\
  &-&{1\over2\pi N}\sum_pK_2(\Lambda-\Lambda_p)
  +{1\over2\pi N}\sum_hK_2(\Lambda-\Lambda_h)\nonumber\\
  &-&{1\over2\pi24N^2\sigma_t(\bp)}K_2^{\prime}(\Lambda-\bp)+
  {1\over2\pi24N^2\sigma_t(\bm)}K_2^{\prime}(\Lambda-\bm)\nonumber\\
  &-&\int\limits_{\bm}^{\bp}K_2(\Lambda-\Lambda^{\prime})\sigma_t(\Lambda')\,.
\eea
where the limits $\bp$ and $\bm$ defined by the equations
\bml\label{bepm}
\be
z(B^{\pm})=\pm{J_F\over N}\,,
\ee
which are equivalent to
\be\label{Bepm}
\int\limits_{\bp}^{\infty}\sigma_t(\Lambda)=\int\limits_{-\infty}^{\bm}
\sigma_t(\lambda)=
{1\over2}\left\{1-n-{n(k)+n(p)-n(h)+2\delta\over N}\right\}\,.
\ee
\eml
As (\ref{sigmat2}) is a linear equation, the density is a sum:
\be
\sigma_t(\Lambda)=\sigma_b(\Lambda)+{1\over N}\sigma_{\{k\}}(\Lambda)+
{1\over N}\sigma_{\{p\}}(\Lambda)+{1\over N}\sigma_{\{h\}}(\Lambda)+
{1\over N^2}\sigma_{fsc}(\Lambda)\,,
\ee
where all terms (the contributions of the bulk, massive particles,
the particles and holes, and the finite size corrections respectively) 
satisfy equations of the type
\be\label{inteq}
x(\Lambda)={\cal I}_x(\Lambda)-{1\over2\pi}\int\limits_{\bm}^{\bp}
  K_2(\Lambda-\Lambda')x(\Lambda')\,,
\ee
where the inhomogeneous part, ${\cal I}_x(\Lambda)$, can be written as:
\bml
\bea
&&\sigma_b(\Lambda):\quad {\cal I}_b(\Lambda)=\sigma_0(\Lambda)\,,\\
&&\sigma_{\{k\}}(\Lambda):\quad {\cal I}_{\{k\}}(\Lambda)=
-{1\over2\pi}\sum_{j=1}^{n(k)}K_1(\Lambda-\sin k_j)\,,\\
&&\sigma_{\{p\}}(\Lambda):\quad {\cal I}_{\{p\}}(\Lambda)=
-{1\over2\pi}\sum_pK_2(\Lambda-\Lambda_p)\,,\\
&&\sigma_{\{h\}}(\Lambda):\quad {\cal I}_{\{h\}}(\Lambda)=
+{1\over2\pi}\sum_hK_2(\Lambda-\Lambda_h)\,,\\
&&\sigma_{fsc}(\Lambda):\quad {\cal I}_{fsc}(\Lambda)=
-{1\over48\pi\sigma_t(\bp)}K_2^{\prime}(\Lambda-\bp)+
 {1\over48\pi\sigma_t(\bm)}K_2^{\prime}(\Lambda-\bm)\,.
\eea
\eml
The equations for the holes and particles can be given in terms of 
the density: noticing, that the particles and holes will condense in the
vicinity of the Fermi points, based on (\ref{Lambdak2}), (\ref{bepm}) and 
(\ref{sigmat0}) we may write:
\bea\label{intsigma}
&&2\pi\int\limits_{\bp}^{\Lambda_{p/h}^+}\sigma_t(\Lambda)={2\pi\over N}
\Delta J_{p/h}^+\,\quad
\left(\Delta J_{p/h}^+=J_{p/h}^+-J_F={\rm half-integer}\right)\,,\nonumber\\
&&2\pi\int\limits^{\bm}_{\Lambda_{p/h}^-}\sigma_t(\Lambda)={2\pi\over N}
\Delta J_{p/h}^-\,\quad
\left(\Delta J_{p/h}^-=-J_F-J_{p/h}^-={\rm half-integer}\right)\,.
\eea
Here $p/h$ is either $p$ or $h$, and the upper index $+$ or $-$ refers to the 
side of Fermi sea to which the particle or hole is near, and the $2\pi$ 
factor is introduced for later convenience.
Equations of the type (\ref{inteq}) in general 
(unlike the half-filled-band case) can not be solved in a 
closed form, neither the Eqs.~(\ref{intsigma}) can be given in a 
more explicit way, nevertheless they can be handled in the $u\to0$ limit.  

\bigskip
\centerline{\it 4. Unbound electrons}
\medskip
The unbound electrons are described by Eqs.\ (\ref{5})
from which it is possible to eliminate the sum over the $\Lambda$s
using the summation formula (\ref{EuMac}) leading to 
\bea\label{unbound2}
 2\pi I_j&=&Nk_j - \sum_{\alpha = 1}^{n(\lambda)}
 \varphi_1(\sin k_j-\lambda_{\alpha})-\nonumber\\
 &-&\sum_{p}\varphi_1(\sin k_j-\Lambda_p)+
 \sum_{h}\varphi_1(\sin k_j-\Lambda_h)-\nonumber\\
 &-&{1\over24N\sigma_t(\bp)}K_1(\Lambda-\bp)+
  {1\over24N\sigma_t(\bm)}K_1(\Lambda-\bm)-\nonumber\\
  &-&N\int\limits_{\bm}^{\bp}
  \varphi_1(\Lambda-\Lambda^{\prime})\sigma_t(\Lambda')\,.
\eea
The integral in (\ref{unbound2}) can be transformed in the following way:
 if a function 
satisfies (\ref{inteq}), than it satisfies also the integral relation
\bea\label{fiegyint}
  \int\limits_{\bm}^{\bp}\varphi_m(\xi-\Lambda)&&\,x(\Lambda)=-
  \left(\int\limits_{-\infty}^{\bm}+\int\limits_{\bp}^{\infty}\right)
  \varphi_m(\xi-\Lambda)\,x(\Lambda)+\nonumber\\
  &&+\int\limits_{-\infty}^{\infty}\varphi_m(\xi-\Lambda)\,{\cal I}_x(\Lambda)-
  \int\limits_{\bm}^{\bp}\varphi_{m+2}(\xi-\Lambda)\,x(\Lambda)\,,
\eea
as it can be checked by calculating the convolution of $\varphi_m$ with 
$x$ of (\ref{inteq}). It is easy to show by iteration, that
this relation is equivalent to
\bea\label{fiegyint2}
  \int\limits_{\bm}^{\bp}&&\varphi_1(\xi-\Lambda)\,x(\Lambda)=\nonumber\\
  &&-\left(\int\limits_{-\infty}^{\bm}+\int\limits_{\bp}^{\infty}\right)
  2\tan^{-1}\tanh{\pi(\xi-\Lambda)\over4u}x(\Lambda)
  +\int\limits_{-\infty}^{\infty}
  2\tan^{-1}\tanh{\pi(\xi-\Lambda)\over4u}
  {\cal I}_x(\Lambda)\,.
\eea
After substituting (\ref{fiegyint2}) into (\ref{unbound2}) and evaluating 
explicitly some of the integrals of the ${\cal I}(\Lambda)$ one arrives at
\bml\label{unbound3}
\bea
 2\pi I_j&=&N\left\{k_j-\int\limits_{-\infty}^{\infty}
 2\tan^{-1}\tanh{\pi(\sin k_j-\Lambda)\over 4u}\sigma_0(\Lambda)\right\}+
 \label{uba}\\
 &+&N\left(\int\limits_{-\infty}^{\bm}+\int\limits_{\bp}^{\infty}\right)
 2\tan^{-1}\tanh{\pi(\sin k_j-\Lambda)\over 4u}\sigma_t(\Lambda)-\label{ubb}\\
 &-&\sum_{\alpha=1}^{n(\lambda)}2\tan^{-1}{\sin k_j-\lambda_{\alpha}\over 4u}
 +\sum_{j'=1}^{n(k)}\phi\left({\sin k_j-\sin k_{j'}\over2u}\right)-\label{ubc}\\
 &-&\sum_p 2\tan^{-1}\tanh{\pi(\sin k_j-\Lambda_p)\over 4u}
 +\sum_h 2\tan^{-1}\tanh{\pi(\sin k_j-\Lambda_h)\over 4u}-\label{ubd}\\
 &-&{1\over24N\sigma_t(\bp)}{\pi\over2u}{1\over\cosh {\pi(\sin k_j-\bp)\over2u}}+
 {1\over24N\sigma_t(\bm)}{\pi\over2u}{1\over\cosh {\pi(\sin k_j-\bm)\over2u}}\,.
 \label{ube}
\eea
\eml
For a general $u$ we find this is the simplest form of the equations for the
unbound electrons, but these equations further simplify in the scaling 
limit.

\bigskip
\centerline{\it 5. The energy and momentum}
\medskip
To calculate the energy, in (\ref{energy}) we evaluate the sum over the 
$\Lambda$ by using again the Euler-Maclaurin formula (\ref{EuMac}):
\bea\label{energy2}
  E=&&-Nut-\sum_j \left( 2t(\cos k_j-u)-\mu\right)
  +\sum_{p}\varepsilon_0(\Lambda_p)
  -\sum_{h}\varepsilon_0(\Lambda_h)-\nonumber\\
  &&-{1\over24 N\sigma_t(\bp)}\varepsilon_0^{\prime}(\bp)+
  {1\over24 N\sigma_t(\bm)}\varepsilon_0^{\prime}(\bm)+
  N\int\limits_{\bm}^{\bp}\varepsilon_0(\Lambda)\sigma_t(\Lambda)\,.
\eea
Iterating $\sigma_t$ in the integral by means of (\ref{sigmat1}) one arrives
at
\bea
E=&&-Nut+N\int\limits_{\bm}^{\bp}\varepsilon(\Lambda)\sigma_0(\Lambda)+
\nonumber\\
  &&+\sum_j \left(-\left( 2t(\cos k_j-u)-\mu\right)-
  {1\over2\pi}\int\limits_{\bm}^{\bp}\varepsilon(\Lambda)K_1(\Lambda-\sin k_j)
  \right)+\nonumber\\
  &&+\sum_{p}\varepsilon(\Lambda_p)
  -\sum_{h}\varepsilon(\Lambda_h)-\nonumber\\
  &&-{1\over24 N\sigma_t(\bp)}\varepsilon^{\prime}(\bp)+
  {1\over24 N\sigma_t(\bm)}\varepsilon^{\prime}(\bm)\,,
\eea
where $\varepsilon(\Lambda)$ satisfies Eq.\ (\ref{inteq}) with 
the inhomogeneous part:
\be
 \varepsilon(\Lambda):\quad
 {\cal I}_{\varepsilon}(\Lambda)=\varepsilon_0(\Lambda)\,.
\ee
(Note, that $\varepsilon\not=\varepsilon_r$ as the limits in the integral
equations (\ref{inteq}) and (\ref{epsilonr}) are different.) 
Next, to calculate the integral of $\varepsilon\times K_1$
we substitute the derivative of (\ref{fiegyint2}), and obtain
\bml\label{energy3}
\bea
E=&&-Nut+N\int\limits_{\bm}^{\bp}\varepsilon(\Lambda)\sigma_0(\Lambda)-
  \label{ena}\\
  &&-\left({1\over24 N\sigma_t(\bp)}\varepsilon^{\prime}(\bp)-
  {1\over24 N\sigma_t(\bm)}\varepsilon^{\prime}(\bm)\right)+\label{enb}\\
  &&+\sum_{p}\varepsilon(\Lambda_p)
  -\sum_{h}\varepsilon(\Lambda_h)+\label{enc}\\
  &&+\sum_j \left(-\left( 2t(\cos k_j-u)-\mu\right)-
  {1\over2\pi}\int\limits_{-\infty}^{\infty}
  {\pi\over2u}{1\over\cosh {\pi(\sin k_j-\Lambda)\over2u}}
  \varepsilon_0(\Lambda)
  \right)+\label{end}\\
  &&+{1\over2\pi}
  \left(\int\limits_{-\infty}^{\bm}+\int\limits_{\bp}^{\infty}\right)
  {\pi\over2u}{1\over\cosh {\pi(\sin k_j-\Lambda)\over2u}}
  \varepsilon(\Lambda)\,.\label{ene}
\eea
\eml
In this expression the first and second terms contain the bulk contributions,
the third term is connected with the finite size corrections (but also the
second term contains finite size corrections due to the deviation of $B^{\pm}$
from $\pm B$), and the rest is the excitation energy.

The momentum is given by (\ref{momentum}), which after summation over
the Fermi sea reads:
\be\label{momentum2}
  P={2\pi\over N}\left(\sum_jI_j-\sum_{\alpha}J_{\alpha}+\sum_{p}J_{p}
  -\sum_{h}J_{h}\right)\,.
\ee
As in (\ref{intsigma}) the particle and hole rapidities are connected to
$\Delta J_{p/h}^{\pm}={\pm}J_{p/h}^{\pm}-J_F$ type expressions, and as 
$J_F$ may have more values, we write
\bea\label{momentum3}
  P=&&\sum_j{2\pi I_j\over N}
  -\sum_{\alpha}{2\pi J_{\alpha}\over N}+
  {\sum_{p}}^{(+)}\,{2\pi\Delta J_{p}^+\over N}-
  {\sum_{h}}^{(+)}\,{2\pi\Delta J_{h}^+\over N}-\nonumber\\
  &&-{\sum_{p}}^{(-)}\,{2\pi\Delta J_{p}^-\over N}+
  {\sum_{h}}^{(-)}\,{2\pi\Delta J_{h}^-\over N}+
  {\pi\delta\over N}(\Delta n^+-\Delta n^-)
  +{\pi N_r\over N}(\Delta n^+-\Delta n^-)\,,
\eea
with
\be
  \Delta n^+=n(p^+)-n(h^+)\,,\quad \Delta n^-=n(p^-)-n(h^-))\,.
\ee
Here the ${\sum}^{(\pm)}$ means summations over the particles and
holes near $\pm J_F$, with $n(p^{\pm})$ resp.\ $n(h^{\pm})$ being the numbers
of the particles resp.\  holes near the Fermi points $\pm J_F$.
As the last term can give an infinite contribution  
in the scaling limit (where the momentum must be devided by $a$), 
we have to redefine the lattice so, that this term be
{\it equivalent} to zero. This can be done, if the bandfilling $n=2N_r/N$ is a 
rational number. Suppose, that $n=l/q$, where $l$ and $q$ are relative 
prime numbers, and that $\nu$ and $\eta$ are the smallest integers satifying
$\nu=4q\eta/l$. If in the redefined lattice $\nu$ lattice sites form one 
elementary cell, than the point $\pi N_r/N=\pi n/2$ of the original
Brillouin zone will be transformed into the origin of the new Brillouin zone,
so the last term of (\ref{momentum3}) can be dropped. (The redefinition of
the lattice has consequences on the parities of some numbers: as the lattice
must consist of an integer number of elementary cells, $N$ and $N_r$ must be 
integer multiples of 4 and 2, respectively)

\section{EQUATIONS OF THE EXCITATIONS IN THE SCALING LIMIT}
\label{sec:eqscl}

\centerline{\it 1. The massless excitations}
\medskip
In Ref.\ \cite{Wo0} the spectrum and structure of the massive excitations 
have been investigated. It has been found, that in order to keep the
mass gap ($m_0$) finite, relations (\ref{scl}) have to be obeyed.
Now we examine the behaviour of the excitations connected with the 
$\Lambda$ distribution in the limit (\ref{scl}).

First consider (\ref{intsigma}). As in the limit (\ref{scl}) 
$N\to\infty$, we divide these equations by $a$:
\be\label{intsigma2}
{2\pi\over a}\int\limits_{\bp}^{\Lambda_{p/h}^+}\sigma_t(\Lambda)=
{2\pi\Delta J_{p/h}^+\over L}\,,\quad\quad
{2\pi\over a}\int\limits^{\bm}_{\Lambda_{p/h}^-}\sigma_t(\Lambda)=
{2\pi\Delta J_{p/h}^-\over L}\,.
\ee   
As the r.h.s.\ is finite, in order to have the l.h.s.\ finite too, 
\be 
\Lambda^{\pm}_{p/h}-B^{\pm}=O(a)\,.
\ee
According to (\ref{scl})
\be\label{a}
   a={4\over\pi m_0}\sqrt{u\sin(\pi n/2)}\,
   {\rm exp}\left\{-{\pi\sin(\pi n/2)\over2u}\right\}\,.
\ee
As the solutions of (\ref{inteq}) change on the scale of $u$, the 
integrands in (\ref{intsigma}) are constants on a scale $\propto a$, 
and we may write:
\be\label{intsigma3}
2\pi\sigma_t(\bp){\Lambda_{p/h}^+-\bp\over a}=
{2\pi\Delta J_{p/h}^+\over L}\,,\quad\quad
2\pi\sigma_t(\bm){\bm-\Lambda_{p/h}^-\over a}=
{2\pi\Delta J_{p/h}^-\over L}\,.
\ee
Finally, as $\sigma_t(B^{\pm})=\sigma_r(B)+O(N^{-1})$, in the scaling limit,
one obtains:
\be\label{intsigma4}
2\pi\sigma_r(B){\Lambda_{p/h}^+-\bp\over a}=
{2\pi\Delta J_{p/h}^+\over L}\,,\quad\quad
2\pi\sigma_r(B){\bm-\Lambda_{p/h}^-\over a}=
{2\pi\Delta J_{p/h}^-\over L}\,.
\ee
These equations can be considered as the secular equations of the 
massless excitations in the scaling limit. They look very similar to
the secular equations of free particles, but they are not exactly 
the same, as $\bp$ and $\bm$, i.e.\ the location of the Fermi 
surface in the rapidity space depend on the number of excitations. 

\bigskip
\centerline{\it 2. Massive particles}
\medskip
Now we consider the scaling limit of (\ref{unbound3}). The first term on 
the r.h.s.\ (line (\ref{uba})) can be evaluated to give
\be
 k_j-\int\limits_{-\infty}^{\infty}
 2\tan^{-1}\tanh{\pi(\sin k_j-\Lambda)\over 4u}\sigma_0(\Lambda)=
 {4\over\pi}\sqrt{u}\exp\left\{-{\pi\over2u}\right\}\,.
\ee
This, in the limit (\ref{scl}) disappears for any $n<1$. 

In the second term (line (\ref{ubb}))
\bea
 N\left(\int\limits_{-\infty}^{\bm}+\int\limits_{\bp}^{\infty}\right)&&
 2\tan^{-1}\tanh{\pi(\sin k_j-\Lambda)\over 4u}\,\sigma_t(\Lambda)=\nonumber\\
 {L\over a}&&
 \left(-\int\limits_{-\infty}^{\bm}+\int\limits_{\bp}^{\infty}\right)
 2\exp\left\{-\left\vert{\pi(\sin k_j-\Lambda)\over 2u}\right\vert\right\}
 \sigma_t(\Lambda)\,.
\eea
As in the limit (\ref{scl}) $B^{\pm}=\pm B+O(1/N)$ (this  can bee seen 
comparing(\ref{Be}) and (\ref{Bepm})) and $\sigma_t=\sigma_r+O(1/N)$,
we may replace $B^{\pm}$ by $\pm B$, resp.\ $\sigma_t$ by $\sigma_r$
(the errors introduced this way are proportinal to $a/u$ resp.\ $a$).
Thus, after some simple manipulations we have
\bea
 N\left(\int\limits_{-\infty}^{\bm}+\int\limits_{\bp}^{\infty}\right)&&
 2\tan^{-1}\tanh{\pi(\sin k_j-\Lambda)\over 4u}\sigma_t(\Lambda)\sim\nonumber\\
 &&L{4\over a}\exp\left\{-{\pi B\over2u}\right\}
 \left(\int\limits_{0}^{\infty}
 \exp\left\{-{\pi\Lambda\over 2u}\right\}
 \sigma_r(B+\Lambda)\right)
 \sinh\left\{{\pi\sin k_j\over2u}\right\}\,.
\eea
Finally using that (\cite{WoPe}, \cite{Wo0})
\be\label{intexpsigma}
\int\limits_{0}^{\infty}
 \exp\left\{-{\pi\Lambda\over 2u}\right\}
 \sigma_r(B+\Lambda)=
 {2u\over\pi}\sqrt{\pi\over e}\sigma(B)\,+
  \,{4u^2\over\pi^2}\sqrt{\pi\over2e}\lim_{u\to0}{\sigma_{0}^{\prime}(B)}
\ee
and 
\be\label{sigmaBlim}
\lim_{u\to0}\sigma(B)={1\over\sqrt{2}}\lim_{u\to0}\sigma_0(B)
\ee
to leading order,
after substituting the value of $B$ (\ref{B(un)}),
we find that the contribution of this term 
to the r.h.s.~of (\ref{5}) is 
\be
Lm_0\sinh\left\{{\pi\sin k_j\over2u}\right\}\,,
\ee 
with $m_0$ given by (\ref{scl}).

The next two terms (line (\ref{ubc}), the contribution of the $\lambda$s and 
$k$s) survives the limit. The contribution of the holes and
particles in the $\Lambda$ distribution (line (\ref{ubd}) tends to a constant
\be
  {\pi\over2}(\Delta n^+-\Delta n^-)\,,
\ee
while the last two terms (the finite size effects, line (\ref{ube})) disappear.

In the parity prescription for $I_j$ there is a parameter, $n(\Lambda)$,
tending
to infinity in the scaling limit. It can be replaced by finite numbers in the
following way: 
first of all, as $n(\Lambda)=n_F+n(p)-n(h)$ with $n_F$ being the number 
of elements in the Fermi sea, and that the total number of `excitations'
$n(k)+n(p)+n(h)=$even, $n(\Lambda)=n_F+n(k)\,({\rm mod}\,2)$.
It is to be noted, that $n_F=N_r+\delta$,
i.e.~it can be both even and odd, but due to the redefinition of the lattice
$N_r$ is always even, thus we have 
\be
I_j={n(\lambda)+n(k)+\delta\over2}{\modegy}\,.
\ee

Collecting all the terms, and introducing the notations  
\be
\kappa_j={\pi\sin k_j\over2u},\quad(n(\kappa)=n(k)),
\quad\quad \chi_{\alpha}={\pi\lambda_{\alpha}\over2u},\quad
(n(\chi)=n(\lambda)),
\ee
we conclude, that the massive particles are described by the equations
\bml\label{massive}
\bea\label{massive.a}
  Lp(\kappa_j)=&&2\pi I_{j}^{\prime}-\sum_{j'}
  \phi\left({\kappa_j-\kappa_{j'}\over\pi}\right)+
  \sum_{\alpha}2{\rm tan}^{-1}\left({\kappa_j-\chi_{\alpha}
  \over\pi/2}\right)\,,\\
  \biggl(I_j^{\prime}=&&I_j-{\Delta n^+-\Delta n^-\over4}=\nonumber\\
  =&&{2n(\chi)+2n(\kappa)+2\delta-(\Delta n^+-\Delta n^-)\over4}\modegy
  \nonumber\\
  =&&\pm{n(\kappa)-2n(\chi)+2m\over4}+{n\over2}\modegy
  \biggr)\,,\nonumber
\eea
\bea\label{massive.b}
  \sum_j 2{\rm tan}^{-1}\left({\chi_{\alpha}-\kappa_j\over\pi/2}\right)=
  &&2\pi J_{\alpha}+
  \sum_{\alpha'}2{\rm tan}^{-1}
  \left({\chi_{\alpha}-\chi_{\alpha'}\over\pi}\right)\,,\\
  \biggl(J_{\alpha}=&&{n(\kappa)+n(\chi)+1\over2}\modegy\biggr)\,,\nonumber
\eea                             
\eml
with 
\be
p(\kappa)=m_0\sinh\kappa
\ee
Eqs.~(\ref{massive}) are the secular equations of the massive particles.
These equations have the same structure as the corresponding equations
of the half-filled band case, with one significant difference:
the parity prescription for the quantum numbers $I_j^{\prime}$ 
depend not only on the number of excitations, but also on the parameter
$\delta$. (For later purposes in the last row of (\ref{massive.a})
we give the $I_j^{\prime}$ in terms of other quantum numbers $n$ and $m$
defined in (\ref{nandm}).)

\section{ENERGY AND MOMENTUM IN THE SCALING LIMIT}
\label{sec:spscl}
\centerline{\it 1. The energy}
\medskip
Now we calculate the scaling limit of (\ref{energy3}). First we calculate
the contribution of the unbound electrons (lines (\ref{end}) and (\ref{ene})).
The expression in (\ref{end}) is exactly the same, as the energy of a 
massive particle in the half-filled band:
\be
-\left( 2t(\cos k-u)-\mu\right)-
  {1\over2\pi}\int\limits_{-\infty}^{\infty}
  {\pi\over2u}{1\over\cosh {\pi(\sin k-\Lambda)\over2u}}
={8t\over\pi}\sqrt{u}\,
   {\rm exp}\left\{-{\pi\over2u}\right\}\,ch\kappa\,.
\ee
These terms disappear in the scaling limit. In (\ref{ene}) we may replace
$B^{\pm}$ by $\pm B$ and $\varepsilon(\Lambda)$ by $\varepsilon_r(\Lambda)$, 
and we arrive at
\bea
{1\over2\pi}
  \left(\int\limits_{-\infty}^{\bm}+\int\limits_{\bp}^{\infty}\right)&&
  {\pi\over2u}{1\over\cosh {\pi(\sin k-\Lambda)\over2u}}
  \varepsilon(\Lambda)=\nonumber\\
 &&={1\over u}\exp\left\{-{\pi B\over2u}\right\}
 \left(\int\limits_{0}^{\infty}
 \exp\left\{-{\pi\Lambda\over 2u}\right\}
 \varepsilon_r(B+\Lambda)\right)
 \cosh\left\{{\pi\sin k\over2u}\right\}\,.
\eea
The integral on the r.h.s.~can be evaluated ({\cite{WoPe},\cite{Wo0}):
\be\label{intexpepsilon}
\int\limits_{0}^{\infty}
 \exp\left\{-{\pi\Lambda\over 2u}\right\}
 \varepsilon_r(B+\Lambda)=
 {4u^2\over\pi^2}\sqrt{\pi\over2e}\lim_{u\to0}{\varepsilon_{0}^{\prime}(B)}\,.
\ee
(The structure of this relation is the same as that of (\ref{intexpsigma}), 
the
difference is due to (\ref{epsB=0}).) Finally we find that
the contribution of the massive particles is simply given by
\be\label{e1}
\sum_j \epsilon(\kappa_j)\,,
\ee
where
\be
\epsilon(\kappa)=m_0\cosh\kappa\,.
\ee

Next we consider the energy contribution of the  particles and holes in the 
$\Lambda$ distribution (\ref{enc}). 
Due to (\ref{epsB=0})
\be
\varepsilon(\Lambda)=\varepsilon_r(\Lambda)+o(B^{\pm}\mp B)\,,
\ee
thus $\varepsilon$ can be replaced by $\varepsilon_r$. Moreover, as the
$\varepsilon_r$ changes on a scale $\propto u$, but 
$\Lambda^{\pm}\mp B\propto a$, we may linearize $\varepsilon_r$
around $\pm B$. Finally we arrive at
\bea
  \sum_{p}\varepsilon(\Lambda_p)
  &&-\sum_{h}\varepsilon(\Lambda_h)=\nonumber\\
  &&=
  {\sum_{p}}^{(+)}\varepsilon_r^{\prime}(B)(\Lambda_p^+-\bp)
  {-\sum_{h}}^{(+)}\varepsilon_r^{\prime}(B)(\Lambda_h^+-\bp)-\nonumber\\
  &&{-\sum_{p}}^{(-)}\varepsilon_r^{\prime}(B)(\Lambda_p^--\bm)
  {+\sum_{h}}^{(-)}\varepsilon_r^{\prime}(B)(\Lambda_h^--\bm)+\nonumber\\
  &&+\varepsilon_r^{\prime}(B)(\Delta n^+(\bp-B)+\Delta n^-(-\bm-B))\,.
\eea
This way the contributions of the particles and holes is expressed by the
quantities known from (\ref{intsigma4}):
\bea\label{e2}
  \sum_{p}\varepsilon(\Lambda_p)
  &&-\sum_{h}\varepsilon(\Lambda_h)=\nonumber\\
  &&=
  {2\pi v\over L}\left\{{\sum_{p}}^{(+)}\Delta J_p^+
  {-\sum_{h}}^{(+)}\Delta J_h^+
  {+\sum_{p}}^{(-)}\Delta J_p^-
  {-\sum_{h}}^{(-)}\Delta J_h^-\right\}+\nonumber\\
  &&+2\pi v
  \left(\Delta n^+{\sigma_r(B)(\bp-B)\over a}
  +\Delta n^-{\sigma_r(B)(-\bm-B)\over a}\right)\,,
\eea
where
\be\label{ve}
v={a\varepsilon_r^{\prime}(B)\over2\pi\sigma_r^{\prime}(B)}\,.
\ee
We note that due to (\ref{epsB=0}), $\varepsilon_r^{\prime}(\Lambda)$
satisfies an equation of the type (\ref{epsilonr}), with
$\varepsilon_0(\Lambda)$ replaced by $\varepsilon_0^{\prime}(\Lambda)$. A 
consequence of this is that analogously to (\ref{sigmaBlim})
\be\label{epsilon'Blim}
\lim_{u\to0}\varepsilon_r^{\prime}(B)={1\over\sqrt{2}}\lim_{u\to0}
\varepsilon_0^{\prime}(B)\,,
\ee
which leads to
\be
v=1
\ee
in the scaling limit.

The terms in line (\ref{enb}), after replacing 
$\varepsilon^{\prime}(B^{\pm})$
and $\sigma_t(B^{\pm})$ by $\varepsilon_r^{\prime}(B)$
and $\sigma_r(B)$, respectively,
turn out to be
\be\label{e3}
-{\pi\over6L}\,.
\ee

Finally we consider the terms in line (\ref{ena}), where the integral is  
obviously a function of $\bm$ and $\bp$ but it also depends 
implicitly on $\varepsilon(\Lambda)$. It can be expanded into a power
series of $(\bp-B)$ and $(\bm+B)$ and after a straightforward calculation 
we find, that
\be\label{e4}
-Nut+N\int\limits_{\bm}^{\bp}\varepsilon(\Lambda)\sigma_0(\Lambda)=
E_r+{1\over2}2\pi L\left\{\left({\sigma_r(B)(\bp-B)\over a}\right)^2+
\left({\sigma_r(B)(\bm+B)\over a}\right)^2\right\}\,.
\ee

Collecting all the terms (\ref{e1}-\ref{e4}) we have
\bea\label{energia}
E=&&E_r-{\pi\over6L}+\nonumber\\
  &&+{1\over2}2\pi L\left\{\left({\sigma_r(B)(\bp-B)\over a}\right)^2+
  \left({\sigma_r(B)(\bm+B)\over a}\right)^2\right\}+\nonumber\\
  &&+2\pi\left(\Delta n^+{\sigma_r(B)(\bp-B)\over a}
  +\Delta n^-{\sigma_r(B)(-\bm-B)\over a}\right)\nonumber\\
  &&{2\pi\over L}\left\{{\sum_{p}}^{(+)}\Delta J_p^+
  {-\sum_{h}}^{(+)}\Delta J_h^+
  {+\sum_{p}}^{(-)}\Delta J_p^-
  {-\sum_{h}}^{(-)}\Delta J_h^-\right\}+\nonumber\\
  &&+\sum_j \epsilon(\kappa_j)\,.
\eea

Next one has to calculate the quantities 
${\sigma_r(B)(\bp-B)/a}$ and ${\sigma_r(B)(-\bm-B)/a}$ and to do this
we use the conditions (\ref{Be}) and (\ref{Bepm}).
Taking the difference of these two equations, and keeping only the terms of 
order of $1/N$, we arrive at the relation
\bea
C\left\{\sigma_r(B)(\bp-B)+{n(p^+)-n(h^+)\over N}\right\}&&
-\sigma_r(B)(\bp-B)=\nonumber\\
=&&-{1\over2N}\left(n(\kappa)+\Delta n^++\Delta n^-+2\delta\right)\,,
\eea
where
\be\label{Ce}
C=\int\limits_B^{\infty}\rho(\Lambda)\,,
\ee
with $\rho$ determined by the equation
\be\label{ro}
\rho(\Lambda)=-{1\over2\pi}K_2(\Lambda-B)-{1\over2\pi}\int\limits_{-B}^{B}
K_2(\Lambda-\Lambda^{\prime})\rho(\Lambda^{\prime})\,,
\ee
and at an analogous relation for $(-B-\bm)$. In the $u\to0$ limit Eq.~(\ref{ro})
can be transformed into a Wiener-Hopf type equation and the integral
(\ref{Ce}) can be calculated:
\be
C=1-\sqrt{2}\,.
\ee
This way we find, that
\be
{\sigma_r(B)(\bp-B)\over a}+{n(p^+)-n(h^+)\over L}=
-{n(\kappa)+3\Delta n^++\Delta n^-+2\delta\over2\sqrt{2}L}\,.
\ee
Substituting this and the analogous expression for $(-B-\bm)$ 
into (\ref{energia}) yields
\bea\label{Energia}
E=&&E_r+\sum_j \epsilon(\kappa_j)-{\pi\over6L}+\nonumber\\
  &&+{2\pi\over L}{1\over2}
  \left\{\left({n(\kappa)+2\Delta n^++2\Delta n^-+2\delta\over2}\right)^2+
  \left({\Delta n^+-\Delta n^-\over2}\right)^2\right\}
  \nonumber\\
  &&+{2\pi\over L}\left\{{\sum_{p}}^{(+)}\Delta J_p^+
  -{\sum_{h}}^{(+)}\Delta J_h^+-{(\Delta n^+)^2\over2}\right\}+\nonumber\\
  &&+{2\pi\over L}\left\{{\sum_{p}}^{(-)}\Delta J_p^-
  -{\sum_{h}}^{(-)}\Delta J_h^--{(\Delta n^-)^2\over2}\right\}\,.
\eea
Note, that the values of the expressions in the last two curly brackets
are always nonnegative integers.

\bigskip
\centerline{\it 2. The momentum}
\medskip
The momentum on a lattice is a dimensionless number, while in a continuum 
it has the dimension $1/length$. In the continuum limit the 
momentum, $P$, is obtained by taking the 
limit of $P_l/a$ (with $P_l$
being the lattice momentum). Now $P_l$ is given by  
(\ref{momentum3}) which after dividing by $a$ and substituting 
Eqs.\ (\ref{massive}) yields:
\bea\label{momentum4}
  P=&&\sum_j p(\kappa_j)+\nonumber\\
  &&+{\sum_{p}}^{(+)}\,{2\pi\Delta J_{p}^+\over L}-
  {\sum_{h}}^{(+)}\,{2\pi\Delta J_{h}^+\over L}-\nonumber\\
  &&-{\sum_{p}}^{(-)}\,{2\pi\Delta J_{p}^-\over L}+
  {\sum_{h}}^{(-)}\,{2\pi\Delta J_{h}^-\over L}+\nonumber\\
  &&+{\pi\over2L}(n(\kappa)+2\delta)(\Delta n^+-\Delta n^-)\,,
\eea
that is
\bea\label{momentum5}
  P=&&\sum_j p(\kappa_j)+{2\pi\over L}
  \left({n(\kappa)+2\Delta n^++2\Delta n^-+2\delta\over2}\right)
  \left({\Delta n^+-\Delta n^-\over2}\right)+\nonumber\\
  &&+{2\pi\over L}\left\{{\sum_{p}}^{(+)}\,\Delta J_{p}^+-
  {\sum_{h}}^{(+)}\,\Delta J_{h}^+-
  {(\Delta n^+)^2\over2}\right\}-\nonumber\\
  &&-{2\pi\over L}\left\{{\sum_{p}}^{(-)}\,\Delta J_{p}^--
  {\sum_{h}}^{(-)}\,\Delta J_{h}^--
  {(\Delta n^-)^2\over2}\right\}\,.
\eea

\section{INTERPRETATION}
\label{sec:interpr}

The secular equations, Eqs.~(\ref{intsigma4}) (for the massless sector) and
(\ref{massive}) (for the massive sector) do not admit 
an immediate interpretation in terms of 
scattering states of a massive and a massless SU(2) doublet 
as this has been the case for the scaling limit of 
the HF Hubbard chain \cite{WoFo}. We present below our interpretation
of the results obtained in the previous sections.  

\begin{itemize}
\item[(i)] The ground state (the physical vacuum) is the lowest energy state
of $N_r$ bound pairs of bare particles. This state can be excited by 
intoducing `particles' and `holes' into the ground-state distribution of the
bound pairs and/or introducing unbound electrons. The former set of excitations
resembles a set of massless particles, while the unbound electrons are
massive. The two sectors do not completely  decouple:
\item[]--- the total number of (massive and massless) excitations must be
even, this is a consequence of the parity prescriptions the quantum numbers
must obey, and the requirement, that the Fermi-sea is symmetric;
\item[]--- the parity prescription for the quantum numbers of the 
massive particles depends on the state of the massless sector;
\item[]--- in the energy the numbers of massive and massless particles
are nonlinearly coupled.
\item[(ii)] As we have noted,
the definition of the Fermi sea we used is only one of several 
(actually infinite) possibilities: the same state can be described in terms 
of a larger or smaller Fermi sea. We may choose $\delta$ even to be a function
of of the described state, but the rule
\be\label{deltapar}
n(\kappa)+n(\Lambda)=n_r(\Lambda)+\delta\modketto
\ee
must be obeyed. It is obvious, that $n(p)$ and $n(h)$ (so $\Delta n^+$
and $\Delta n^-$) depend on the definition
of the Fermi sea, but the quantity entering the energy and the momentum
\be\label{devi}
  n(\kappa)+2\Delta n^++2\Delta n^-+2\delta=N_e-2N_r
\ee
is independent of the definition of $\delta$, as it gives
the deviation of the actual bare particle number from 
that of the reference state. 
It is easy to see, that the other quantity which enters into 
the energy and momentum
\be
\Delta n^+-\Delta n^-
\ee
is also independent of $\delta$, hence no physical (measurable)
quantity depends
on our choice of the Fermi sea. 
It is, however, unavoidable to make a definite choice in order to 
be able to define the massless particles. 

\item[] A simple consequence of the above arbitrariness
is that
we can formally `minimize' the coupling between the massive and massless
sectors: we can choose a $\delta$ (obeying (\ref{deltapar})) such 
that 
$\delta'=n(\kappa)/2+\delta$ have the smallest modulus. There are
four inequivalent values for $\delta'$: 0 or 1, if $n(\kappa)$ is even, 
and +1/2 or -1/2, if $n(\kappa)$ is odd. We note, that Eq.\ (\ref{devi})
has now the form
\be\label{devi'}
  2\Delta n^{+\prime}+2\Delta n^{-\prime}+2\delta'=N_e-2N_r\,.
\ee
The `minimizing' of the coupling between the massive and massless
sectors corresponds to choosing
the Fermi level with
$2\Delta n^{+\prime}+2\Delta n^{-\prime}$ as close to $N_e-2N_r$ as possible.
It is worth to mention that in the HF band case $2N_r-N_e=N-N_e$
is measured by the number of massless particles only, 
in this sense the above choice of
$\delta'$ mimicks most the HF case.

\item[(iii)] To specify a state it is not sufficient to give the
number of particles (and the quantum numbers) in the two sectors, but
one also needs the value of $\delta$ or $\delta'$. 
It is not hard to see,
that any state with given $n(\kappa)$, $n(p)$ and $n(h)$ can have both
$\delta=0$ and $\delta=1$ (if we use the convention (\ref{delta})), 
or equivalently can have both $\delta'=0$ and $\delta'=1$ for $n(\kappa)$ 
even, 
and $\delta'=\pm1/2$ for $n(\kappa)$ odd. 
In this sense $\delta$ (or $\delta'$) is a free parameter. 
States differing only in the value of $\delta$ ($\delta'$)
are not degenerate, as $\delta$ appears
in the parity prescription for $I_j^{\prime}$ in (\ref{massive.a}), and
it appears explicitly in the energy (\ref{Energia}) too. 

\item[(iv)] The massive sector described by 
Eqs.~(\ref{massive}) consists of (relativistic) particles
in the doublet representation of SU(2). 
Their contribution to the energy and momentum is 
not simply the sum of the individual
contributions
but there are terms both in the energy and
momentum, which depend on the {\it number of massive particles} in a 
nontrivial way. (As these terms contain data on the massless sector too,
we interpret them as a coupling between the two sectors.)
\item[] Any solution of Eqs.~(\ref{massive}) 
corresponds to a highest weight state of SU(2) with $S^2=l(l+1)$,
$S^z=m$, $l=m=(n(\kappa)-2n(\chi))/2$. All the other members of the 
$l=(n(\kappa)-2n(\chi))/2$ multiplets are degenerate with this state, and
can be obtained by the action of the 
$\sigma^-=\sum_i c_{i,\downarrow}^+c_{i,\uparrow}$ operator.
\item[] The BA Eqs.~(\ref{massive}) are of the familiar type
and this makes possible extract
the two particle scattering matrices in the usual way
\cite{AnLo2,Kor,DeLo1}.
For this we have to choose those solutions of Eqs.~(\ref{massive})
which correspond to the triplet and singlet states with an 
{\it empty massless sector}. 
To avoid ambiguities due to the freedom in choosing
the Fermi level one should first define
the state with no massless particles. It is natural to consider
the massless sector empty, if its energy contribution is zero. This 
corresponds to $m=n=0$ (see the next point), and then the
scattering matrix of the massive doublet is given by
\be\label{smm}
\hat S(\Delta\kappa)=-\exp\left\{i\phi\left({\Delta\kappa\over\pi}\right)
\right\}
{\Delta\kappa{\hat{\rm I}}-i\pi{\hat \Pi}\over \Delta\kappa-i\pi}\,,
\ee
where $\hat{\rm I}$ resp.\  $\hat \Pi$ are the identity resp.\ permutation operators 
acting on the spins of the two particles:
\be
{\rm I}^{\sigma'_1\sigma'_2}_{\sigma_1\sigma_2}=
\delta_{\sigma'_1\sigma^{\phantom{'}}_1}
\delta_{\sigma'_2\sigma^{\phantom{'}}_2}\ ,\quad
\Pi^{\sigma'_1\sigma'_2}_{\sigma_1\sigma_2}=
\delta_{\sigma'_1\sigma^{\phantom{'}}_2}
\delta_{\sigma'_2\sigma^{\phantom{'}}_1}\ .
\ee

\item[(v)] 
The massless sector described by Eqs.\ (\ref{intsigma4}). 
looks very much like as if it consisted of free 
particles. This picture, however, cannot be taken at face value as the 
definition of the particles (and holes) is not unique. Also as both the
energy and momentum contain the {\it collective} terms it looks
as if there were interactions. One cannot, however, extract phaseshifts
from Eqs.\ (\ref{intsigma4}).
\item[] The energy and momentum can be described in terms of a 
CFT. If the massive sector is empty the 
energy momentum dispersion shows a tower structure: the central 
charge is $c=1$, and the apexes of the towers are located at 
\be
E-E_r=-{\pi\over6L}+{2\pi\over L}x_{n,m}\,,\quad P={2\pi\over L}s_{n,m}\,,
\ee
with
\be
x_{n,m}={1\over2}\left(n^2+m^2\right)\,,\quad s_{n,m}=nm\,,
\ee
where the integers $n$ and $m$ are
\be
n=\left({\Delta n^++\Delta n^-+\delta}\right)\,,\quad 
m=\left({\Delta n^+-\Delta n^-\over2}\right)\,.
\ee
This leads to the conformal weights (the notations are those of \cite{Car})   
\be\label{sulyok}
\Delta={1\over4}(n+m)^2\,,\quad\bar\Delta={1\over4}(n-m)^2\,.
\ee
(We have to note, that even if the massive sector is not empty, the 
contribution of the massless sector is of the tower structure, just with
\be\label{nandm}
n=\left({n(\kappa)+2\Delta n^++2\Delta n^-+2\delta\over2}\right)\,,\quad 
m=\left({\Delta n^+-\Delta n^-\over2}\right)\,.
\ee
In this case, however, due to the coupling of the massive sector, 
the apexes of the 
towers can not be interpreted as anomalous dimensions and spins.
For $n(\kappa)\not=0$ $n$ and $m$ can be half-integers too:
$2n,2m=n(\kappa)\modegy$.)
The conformal weights (\ref{sulyok}) correspond to a SU(2)$\times$SU(2)
symmetric CFT. 
It should be interesting to directly identify the
corresponding SU(2)$\times$SU(2) multiplet structure.

\item[(vi)] The spectrum of the limiting model
is the same  both in the NHF and the HF band case.
Actually the complete energy and momentum are
\be\label{cenergia}
  E=E_r+\sum_j \epsilon(\kappa_j)-{\pi\over6L}
  +{2\pi\over L}
  \left({n^2+m^2\over2}\right)+{2\pi\over L}\left(\nu^++\nu^-\right)\,,
\ee
\be\label{cmomentum}
  P=\sum_j p(\kappa_j)+{2\pi\over L}nm+{2\pi\over L}\left(\nu^+-\nu^-\right)\,.
\ee
The towers are obviously the same as in the HF case, and
so is the contribution
of the massive sector, as the parity prescription for the 
quantum numbers in (\ref{massive}) (i.e.\ the quantization of the momenta
of the massive particles) when expressed in terms of $n$ and $m$ agrees with
that of the HF case. 
We note, however, that this way we have only shown
 that the points of the two spectra coincide, but we
can not say anything about the degeneracies of the single points.
\end{itemize}

Finally let us briefly present the results for the ground state energy
of a finite density state (i.e.\ when the number of excitations
is macroscopic) of the limiting model. 
As $\delta$ can be
choosen to minimize the coupling between the sectors, one can discuss 
the massive and massless sectors separately.
(The coupling due to the parity-prescription for $I_j^{\prime}$ 
and the explicit presence of $\delta'$ in the energy of the 
massless sector may actually give a 
contribution $O(1/L)$ in the energy density.)
Since then the massive sector is described by the same equations 
and energy-momentum
dispersion as in the HF-band case, one can 
literally take over the results
obtained there \cite{WoFo} and we do not reproduce the formulae here.

The treatment of the massless sector is extremely simple.
After choosing the $\delta'$ as described in the previous section, the energy
density associated with the massless sector is just given as
\be\label{density}
{\bar{\cal E}}(\varrho)={\pi\varrho^2}\,,
\ee
where $\varrho$ is the density of massless particles.
Eq.\ (\ref{density}) coincides with the energy density of noninteracting
particles.

The total free energy density (defined as
the Legendre transformation of ${{\cal E}}(\varrho)$ with
a chemical potential $\nu$) will be simply the sum of the
massive and massless contributions just as in the HF case.
The decoupling of the two sectors depends apparently on our choice of 
the Fermi level.
We can understand this as follows. There are two energy contributions
controled by two independent parameters.
One is the energy of the massive particles, related to their relativistic
dispersion, this is
determined by $n(\kappa)$. 
The other 
contribution is a (quadratic) function of $N_e-2N_r$, i.e.\ the
deviation of the actual particle number from that of the reference
state.
If we make the Legendre transformation in this two variables,
we shall get two independent terms. If we choose some combinations
of $n(\kappa)$ and $N_e-2N_r$
as independent variables, the free energy will have
terms depending on both chemical potentials. 
(In the HF-band 
case $2N_r-N_e=N-N_e$ is measured by the number of massless particles,
thus the separation of the two sectors is obvious. For the NHF-band
case measuring $N_e-2N_r$ can be done in different ways. After
having choosen the $\delta'$, 
$N_e-2N_r\sim2L\varrho$. The factor 2 in this equation 
appears as $\varrho$ is the density of bound {\em pairs} and this is
the reason for the discrepancy 
of (\ref{density}) from the analogous expression in the HF case.)

In conlusion the above result provide strong  
(although indirect) evidence that the scaling limit
of the NHF Hubbard chain is the SO(4) symmetric CGN model.
\goodbreak
\section{DIFFERENCES BETWEEN THE HF AND NHF CASES}
\label{sec:diffs}

In the formulae (\ref{scl}) one can take the 
$n\to1$ limit suggesting that there is a smooth limit
from the NHF to the HF case.
In this section we list some differences between the HF and NHF
cases, which clearly show, that in spite of the `continuity' of 
(\ref{scl}) in $n$ 
the scaling limit of the NHF and that of the 
HF case can not be related in a trivial way 
(in other words the $n\to1$ and the scaling limit do not commute in a 
obvious way).
\begin{itemize}
\item In the HF case the ground state is well defined, and in the 
description of the bound pairs ($\Lambda$) the Fermi surface plays no role:
the Fermi points in the rapidity space ($B^{\pm}$) are at infinity.
As a consequence, there are no particles, only holes connected with the
$\Lambda$ distribution. These elementary excitations (we call them 
massless particles) carry an SU(2) isospin and the states are isospin 
eigenstates often characterized by a set of complex variables.
In the NHF case the Fermi surface is at some finite rapidity, there are
both particles and holes, but in the limiting theory there are no states 
with complex $\Lambda$s, as those require an infinite excitation energy. 

\item In the HF case the two kinds of excitations have two different
internal degrees of freedom (spin resp.\  isospin), and the
ground state is a singlet of both. In the NHF case one of the excitations 
carries the spin, but the isospin of the state cannot be uniquely attached
to one or the other kind of excitation. The reference state
is a spin singlet, but its isospin $(N-2N_r)/2\to\infty$ in the scaling
limit. (The difference between the isospin of an excited state and that
of the reference state is, however, finite: 
$\Delta I_3=-(n(\kappa)+2\Delta n^++2\Delta n^-+2\delta)/2$) 

\item In the HF chain the massless particles are uniquely defined, in
the NHF case --- although the state can be given uniquely --- the definition
of the particles and holes connected with the $\Lambda$ distribution 
(massless particles) is not unique.

\item In both cases the conformal dimensions of the massless sector
correspond to an SU(2)$\times$SU(2) symmetric CFT. 
In the case of the HF chain 
an SU(2) symmetry corresponding to the gapless excitations is already 
present in the lattice model, and this with the separation of the left
and right sectors develops into an SU(2)$\times$SU(2) symmetry.
For the NHF chain in the lattice model
there is no such SU(2) symmetry, 
which could be identified in the massless sector
and the SU(2)$\times$SU(2) symmetry (whose existence is indicated
by the conformal weights) appears  only in the scaling limit.

\item While for the HF case the conformal weights $\Delta$ and $\bar\Delta$
are connected to the right resp.\ left going massless particles, for the NHF
case both $\Delta$ and $\bar\Delta$ depend on both the right (+) and left
($-$) part of the massless sector.
\end{itemize}
\appendix

\section{}\label{sec:nlim}

We sketch here briefly the `naive' continuum limit of the less than half 
filled Hubbard chain.
A more detailed description of the procedure is given in Ref.\ \cite{WoEcTr}.

First we write the (\ref{1}) Hamiltonian in the form
\bea\label{Hn}
\hat H=&-&t\sum_{i=1}^N\sum_{\sigma=\uparrow,\downarrow}
  \left( c^{+}_{i,\sigma}c^{\phantom{+}}_{i+1,\sigma}+h.c.\right) +
  {U\over2}\sum_{i=1}^N
  \left( \hat n_{i,\uparrow}+\hat n_{i,\downarrow}\right)^2\nonumber\\
  &+&h\sum_{i=1}^N\left(\hat n_{i,\uparrow} + \hat n_{i,\downarrow}\right)
  +{NU\over4}\,,
\eea
with $h=\mu-U$.

We have seen earlier, that in order to avoid divergent terms in the 
momentum we have to redefine the lattice so, that in the new lattice 
$\nu$ sites form one elementary cell. ($\nu$ together with an other 
number $\eta$ is defined so, that
if the bandfilling is $n=l/q$, where $l$ and $q$ are relative 
prime numbers, than $\nu$ and $\eta$ are the smallest integers satifying
$\nu=4q\eta/l$.)
We define the operators
\be
\phi_{\alpha,\sigma}(n)={1\over\nu\sqrt{a}}\sum_{j=1}^\nu e^{-ik\alpha j}
c_{\nu(n-1)+j,\sigma}\quad \alpha=1,2,\ldots,\nu,\quad
\sigma=\uparrow,\downarrow,
\quad k={2\pi\over\nu}.\ee
In terms of these the original Fermion operators are
\be\label{cek}
c_{\nu(n-1)+j,\sigma}=\sqrt{a}\sum_{\alpha=1}^\nu e^{ik\alpha j}
\phi_{\alpha,\sigma}(n)
\quad j=1,2,\ldots,\nu.
\ee
The part of the Hamiltonian quadratic in the field operators now reads
\bea
&&\sum_{n=1}^{N'}\sum_{\sigma}\sum_{\alpha=1}^{\nu}\nu a(h-2t\cos k\alpha)
\phi_{\alpha,\sigma}^+(n)\phi_{\alpha,\sigma}(n)\nonumber\\
&&-at\sum_{n=1}^{N'}\sum_{\sigma}\sum_{\alpha,\beta=1}^{\nu}
\phi_{\alpha,\sigma}^+(n)
\left(\phi_{\beta,\sigma}(n+1)-\phi_{\beta,\sigma}(n)\right)e^{ik\beta}
\nonumber\\
&&+at\sum_{n=1}^{N'}\sum_{\sigma}\sum_{\alpha,\beta=1}^{\nu}
\phi_{\alpha,\sigma}^+(n)
\left(\phi_{\beta,\sigma}(n)-\phi_{\beta,\sigma}(n-1)\right)e^{-ik\alpha},
\eea
with $N'=N/\nu$ being the number of new elementary cells.
This expression shows that the procedure leads to a 
meaningful result only, if for some $\alpha=\alpha_0$ and
$\alpha=\nu-\alpha_0$
\be\label{alpha0}
h-2t\cos k\alpha=0,
\ee
otherwise all the new fields become infinitely massive if $t\to\infty$.
Obviously
\be
{2\pi\alpha_0\over\nu}=k_F(={\pi n\over2})
\ee
what implies, that
\be
\alpha_0=\eta\,,
\ee 
with $\eta$ defined above. It is clear, that all the modes $\alpha\not=
\eta,\nu-\eta$ can be omitted, as they become infinitely massive in 
the continuum limit. 

The length of the chain is 
\be
L=Na=\nu aN'\ (N'=int.)
\ee 
and the continuum limit is defined as
$$a\to0,\quad N'\to\infty,\quad L=fixed$$
with the continuous variable 
\be
x=\nu a(n-1/2),\quad dx=\nu a.
\ee
In this limit
\be\label{sumint}
\sum_n^{N'}\to\int\limits_0^L{dx\over\nu a}
\quad {\rm and}\quad\delta_{n,n'}\to\nu a\delta(x-x').
\ee
If we introduce
\be\label{pszik}
\phi_{\eta,\sigma}(n)=\psi_{1,\sigma}(x),
\quad\phi_{\nu-\eta,\sigma}(n)=\psi_{2,\sigma}(x),
\ee
then
\be
\left\{\psi_{\alpha,\sigma}(x),\psi_{\beta,\sigma'}^{+}(x')\right\}=
\delta(x-x')\delta_{\alpha,\beta}\delta_{\sigma,\sigma'}.
\ee
Applying (\ref{cek}), (\ref{pszik}) and (\ref{sumint}) to the Hubbard 
Hamiltonian (\ref{Hn}) 
one obtains in the naive continuum limit the following Hamiltonian density:
\bea
{\cal H}(x)=&&(2at\sin\pi n/2)\left(-\sum_{\sigma=1}^2\psi_{\sigma}^+
\gamma_5\pa_x\psi_{\sigma}+\right.\\
&&+\left.u\left[\sum_{\sigma\sigma'}\left(\psi_{1,\sigma}^+\psi_{2,\sigma}
\psi_{2,\sigma'}^+\psi_{1,\sigma'}+
\psi_{2,\sigma}^+\psi_{1,\sigma}
\psi_{1,\sigma'}^+\psi_{2,\sigma'}\right)
+\left(\sum_\sigma\psi_{\sigma}^+\psi_{\sigma}\right)^2\right]\right)
\eea
where
\be
\psi_{\sigma}=\pmatrix{\psi_{1,\sigma}\cr
\psi_{2,\sigma}\cr}\,,\qquad
\gamma_5=\pmatrix{i&0\cr            
                  0&-i\cr}\,,
\ee                   
and $u=U/4t\sin\pi n/2$. (Note that this differs from the definition used in the 
bulk of the paper as there $u=|U|/4t$.)


\newpage
\begin{references}
%
\bibitem{Sha}
B.S.~Shastry, J.~Stat.~Phys.~{\bf50} (1988) 57
%
\bibitem{LiWu}
E.H.~Lieb and F.Y.~Wu, Phys.~Rev.~Lett.~{\bf20} (1968) 1445
%
\bibitem{KoEss}
V.E.~Korepin, F.H.L.~E\ss ler,  
Exactly solvable models of strongly correlated electrons, World Scientific,
Singapore (1994)
%
\bibitem{So}
J.~S\'olyom, Advances in Physics {\bf 28} (1979) p.201
%
\bibitem{GrNe}
D.~Gross and A.~Neveu, Phys.~Rev.~D {\bf 10} (1974) 3235
%
\bibitem{WiLa}
P.B.~Wiegmann and A.I. Larkin, Sov.~Phys.~JEPT {\bf 45} (1977) 448
%
\bibitem{AnLo}
N. Andrei, J.H.~Lowenstein, Phys.~Rev.~Lett.~{\bf43} (1979) 1698
%
\bibitem{Fi}
V.M.~Filev, Teor.i Mat.Fiz,{\bf 33} (1977) 119
%
\bibitem{Me}
E.~Melzer, Nucl.~Phys.~B443[FS] (1995) 553
%
\bibitem{WoFo}
F.~Woynarovich, P.~Forg\'acs, Nucl.~Phys.~B 498 [FS] (1997) 565
%
\bibitem{Wo0}
F.~Woynarovich, J.~Phys.~A {\bf29} (1996) L37
%
\bibitem{Wo1}
F.~Woynarovich, J.~Phys.~C {\bf 15} (1982) 85
%
\bibitem{Wo2}
F.~Woynarovich, J.~Phys.~C {\bf 16} (1983) 6593
%
\bibitem{WoPe}
F.~Woynarovich and K.~Penc, Z.~Phys.~B {\bf 85} (1991) 269
%
\bibitem{KrOv}
V.Ya.~Krivnov and A.A.~Ovchinnikov, Sov.~Physics JETP {\bf 40} (1975) 781 
%
\bibitem{YaYa}
C.N.~Yang, C.P.~Yang, Phys.~Rev.~{\bf 150} (1966) 329
%
\bibitem{AnLo2}
N. Andrei, J.H.~Lowenstein, Phys.~Lett.~{\bf91B} (1980) 401
%
\bibitem{Kor}
V.E.~Korepin; Theor.~Math.~Phys.~76 (1980) 165
%
\bibitem{DeLo1} 
C.~Destri, J.H.~Lowenstein; Nucl.Phys.B205[FS5] (1982) 369
%
\bibitem{Car}
Phase Transitions and Critical Phenomena, ed.\ C.Domb, J.L.Lebowitz;
Academic, New York, 1987.
%
\bibitem{WoEcTr}
F.~Woynarovich, H-P.~Eckle, T.T.~Truong; J.Phys.A {\bf 22} (1989) 4027
%
%
%
%
\end{references}
\end{document}